\documentclass[a4paper,11pt]{article}
\pdfoutput=1 
\usepackage{jinstpub} 
\usepackage{tikz}

\title{CAKE: The Coincidence Array for K600 Experiments}

\author[a,b,1]{P. Adsley,\note{Corresponding author. Present address: Institut de Physique Nucl\'{e}aire d'Orsay, UMR8608, IN2P3-CNRS, Universit\'{e} Paris Sud 11, 91406 Orsay, France}}

\affiliation[a]{Department of Physics, University of Stellenbosch, South Africa}
\affiliation[b]{iThemba LABS, South Africa}
\affiliation[c]{Department of Physics, University of York, Heslington, York, UK, YO10 5DD}
\affiliation[d]{Department of Physics, University of Notre Dame, Notre Dame, Indiana 46556, USA}
\affiliation[e]{School of Physics, University of the Witswatersrand, Johannesburg 2050, South Africa}
\affiliation[f]{Department of Physics, University of the Western Cape, P/B X17, Bellville 7535, South Africa}

\author[b,2]{R. Neveling,\note{Corresponding author.}}
\author[a,b]{P. Papka,}
\author[b]{Z. Dyers,}
\author[a,b]{J.W. Br\"{u}mmer,}
\author[c]{C.Aa. Diget,}
\author[c]{N.J. Hubbard,}
\author[a,b]{K.C.W. Li,}
\author[d,3]{A. Long,\note{Present address: Physics Division, Los Alamos National Laboratory, Los Alamos, New Mexico 87545, USA}}
\author[b,f]{D.J. Marin-Lambarri,}
\author[b,e]{L. Pellegri,}
\author[b,f]{V. Pesudo,}
\author[b]{L.C. Pool,}
\author[b]{F.D. Smit,}
\author[f]{S. Triambak}

\emailAdd{padsley@gmail.com}
\emailAdd{neveling@tlabs.ac.za}

\abstract{The combination of a magnetic spectrometer and ancillary detectors such as silicon detectors is a powerful tool for the study of nuclear reactions and nuclear structure. This paper discusses the recently commissioned silicon array called the \textquoteleft CAKE\textquoteright\ which is designed for use with the K600 magnetic spectrometer at iThemba LABS.}


\arxivnumber{1610.08053} 

\begin{document}
\maketitle
\flushbottom

\section{Introduction}
\label{sec:intro}

The Coincidence Array for K600 Experiments (CAKE) is a new silicon array designed for use with the K600 magnetic spectrometer at iThemba LABS \cite{Neveling}. It enables coincidence spectroscopy of charged-particle decays following inelastic scattering and transfer reactions detected by the focal-plane detector system of the magnetic spectrometer. While scattering and transfer inclusive reactions using hadronic beams and magnetic spectrometers have a long and distinguished history in the study of the properties of nuclei, in these reactions no information is gained concerning the decay of the populated states. The addition of a silicon detector array means that it is possible to observe the charged-particle decays of populated states, which can provide valuable information about the underlying structure of these states.

For example, in recent years, candidate $\alpha$-particle cluster and Bose-Einstein condensate states in light nuclei have been proposed (see Refs. \cite{KCWL_MSc,NevelingVarenna,KCWL_arXiv} and references therein). The properties of these states including the total and partial widths (and therefore branching ratios) can be predicted from various structural models. The combination of the K600 and the CAKE enables detailed spectroscopy of these states to be carried out, providing stringent tests of the predicted nature of these states.

In nuclear astrophysics, many reactions of interest cannot yet be measured in the laboratory due to the low cross sections involved, the instability of one (or more) of the species involved and, in some cases, the requirement for a gaseous target. In the absence of direct measurements of these reactions, the reaction rate can be calculated if detailed knowledge of the properties of the states in the compound nucleus is available. Valuable data on the number and excitation energies of resonances have been gathered using magnetic spectrometers (e.g. \cite{Matic_22Mg}). However, in many cases, information on spin-parities, branching ratios and partial widths are not available. Extending previous studies such as those described in Ref. \cite{Matic_22Mg} to include observation of the charged-particle decays is one way to determine the properties of these states, and thus to constrain the astrophysical reaction rates.

For both the nuclear structure and astrophysics experiments, it is obviously necessary that coincidence measurements are carried out with an efficient array to allow the observation of weak decay branches. Measurements that cover a large angular range with a good angular resolution are also extremely helpful in assigning angular momentum values; of course, an array which covers a large angular range is likely to be highly efficient. 

The use of a magnetic spectrometer for these types of coincidence experiments has an additional benefit. The dispersion of the spectrometer physically separates the unreacted beam from the reaction products of interest. If two silicon arrays are used to probe the $(p,t)$ reaction, as in Ref. \cite{PhysRevC.82.045803}, then the silicon array being used to detect the tritons will also be exposed to scattered protons from the beam. This limits the maximum beam current which may be used (1-2 nA is the example given in Ref. \cite{PhysRevC.82.045803}) with an angle-dependent energy resolution of 65-115 keV. As is discussed in Section \ref{sec:K600}, much higher beam currents are possible with the K600 in $(p,t)$ reactions, and energy resolutions of down to 30 keV are achievable.

\subsection{Requirements of the scientific programme}

The experimental considerations of the scientific programme lead to a number of requirements for the CAKE:

\begin{enumerate}
 \item The silicon detectors comprising CAKE should have a uniform behaviour and threshold without large position-dependent variations in dead-layer thickness and signal size.
 \item The CAKE should cover a wide angular range with a fine segmentation for the polar angle ($\theta$) - doing so allows the angular distribution to be measured over a wide angular region. This distribution is vital in being able to assign orbital angular momentum values for decays, and therefore to infer $J^\pi$ values to decaying states, as well as in determination of the branching ratio where the coincidence yield must be corrected for the missing solid angle.
 \item It must be able to discriminate between decay particles of different species such as protons and $\alpha$ particles.
 \item It must be able to deal with high rates due to beam currents of up to 20 pnA and relatively thick targets (up to around 1 mg/cm$^2$) used in K600 experiments.
 \item It must be designed such that it does not interfere with the operation of the K600 spectrometer, such as inducing beam halo which would be observed on the focal plane.
 \item Setting the beam up to the K600, especially in 0-degree mode, is a time-consuming process (typically lasting around one eight-hour shift) which requires viewing the beam upstream of the scattering chamber and at the target position using thick zinc sulphide beam viewers. It is beneficial - though not mandatory - to perform this tuning without the CAKE in place within the scattering chamber.  The CAKE and the associated scattering chamber should therefore allow for simple and convenient installation and removal of the array so that the beam can be tuned without the CAKE in position before it is installed into the chamber.
\end{enumerate}

The design and operation of the CAKE have been carried out with these requirements in mind, as set out in Section \ref{sec:exp}. However, in order to understand how the CAKE operates with the K600 the spectrometer must briefly be described.

\section{The K600 Magnetic Spectrometer}
\label{sec:K600}

The K600 \cite{Neveling} is a QDD magnetic spectrometer located at iThemba LABS, Cape Town, South Africa consisting of two dipole bending magnets, two trim coils and a quadrupole at the spectrometer entrance. The spectrometer has a solid angle of around 3.5 msr (the exact solid angle depends on the choice of the collimator for the experiment), corresponding to an angular bite of around $\pm2$ degrees. It is possible to determine the scattering angle with a resolution of 0.5 degrees by considering the focal plane trajectory and position (see Refs. \cite{Neveling} and \cite{Si28Paper}). The spectrometer may be used with the aperture centre at angles of: 0, 4, >7 degrees. This means that the entire angular region can be covered with the K600, albeit with different beam stop locations in each case. In the 4-degree (small-angle) mode of operation \cite{Si28Paper}, the beam is stopped before the entrance quadrupole for the spectrometer. At present, higher-angle measurements with the K600 can only be carried out with the older sliding seal scattering chamber and not with the new scattering chamber (see section \ref{sec:AMI} for details) which houses the CAKE. In these measurements, the beam is stopped within the scattering chamber, which would result in a very high background in the silicon detectors, making coincidence experiments impractical.

The K600 has three different focal planes: high- ($p_{\text{max}}/p_{\text{min}} = 1.048$), medium- ($p_{\text{max}}/p_{\text{min}} = 1.097$) and low-dispersion (not used). For inelastic scattering reactions at 0 degrees, the high-dispersion focal plane is used. In this mode of operation, the unscattered beam passes through the spectrometer and past the high-momentum end of the high-dispersion focal plane. The beam is then stopped in the wall of the spectrometer vault. The beam intensity that may be used in these experiments is limited to around 2 pnA due to the  background which otherwise results on the focal plane. The background rate and the signal-to-background ratio are both strongly dependent on the quality of the beam tune and the target species. Targets with a high atomic number produce a higher level of background on the focal plane which may result in a decrease in the maximum practical beam current. The excitation region covered depends on the beam energy and species. For 200-MeV protons and $\alpha$ particles, the excitation region covered in the experiment is from around 9-24 MeV.

The energy resolution in the inelastic scattering experiments is highly dependent on the beam energy and species, and the target thickness. However, as a guide, 200-MeV protons can be detected with resolutions of around 40 keV \cite{LMDarXiv}, FWHM while 200-MeV $\alpha$-particles have resolutions of around 70 keV, FWHM \cite{Si28Paper}.

In ($p,t$), ($^3$He,$t$) and ($^3$He,$d$) reactions, the medium-dispersion focal plane is used. In these cases, the unreacted beam has a lower rigidity than the species of interest and is thus bent more easily by the magnetic field. The beamstop is located inside the first dipole of the K600. In this case, the beam current which can be used is higher (up to around 20 pnA) without the rate in the focal plane becoming too high. However, in this case an additional limitation related to the accelerator system may become a factor. If the slits in the high-energy line coming from the cyclotron are opened in order to increase the beam current, the energy resolution of the focal plane spectrum can worsen.

The focal plane detector suite for the K600 consists of drift chambers for position determination and plastic scintillating paddles which act as the experimental trigger and measure the energy lost by a particle travelling through them. The trigger signal from the paddles is not only used to start the data acquisition but is also used as a reference time for each of the CAEN V1190A time-to-digital converters (TDCs). This ensures that the TDC modules are aligned with one another. Additionally, the time of the next RF pulse from the accelerator control is recorded in the TDCs. The time between the trigger pulse and the next RF pulse is effectively a relative measurement of the time-of-flight through the spectrometer which is used for particle identification.

The operation of the focal-plane drift chambers has been described in detail elsewhere \cite{Neveling} and is not relevant to the current discussion.

Increasing the time separation between beam pulses can result in cleaner focal-plane particle identification and higher quality coincidence data. For this purpose, the accelerator system contains a pulse selector, an RF component located before the cyclotron which deflects a proportion of pulses. The result of this is that the time separation between beam pulses is increased, improving the time-of-flight separation between different species at the focal plane and, in the case of coincidence measurements, allowing clearer correlations between focal-plane and silicon events to be made.

\section{Design and implementation of the CAKE}
\label{sec:exp}

\subsection{The silicon detectors}

\begin{figure}
\centering
\includegraphics[width=0.7\textwidth]{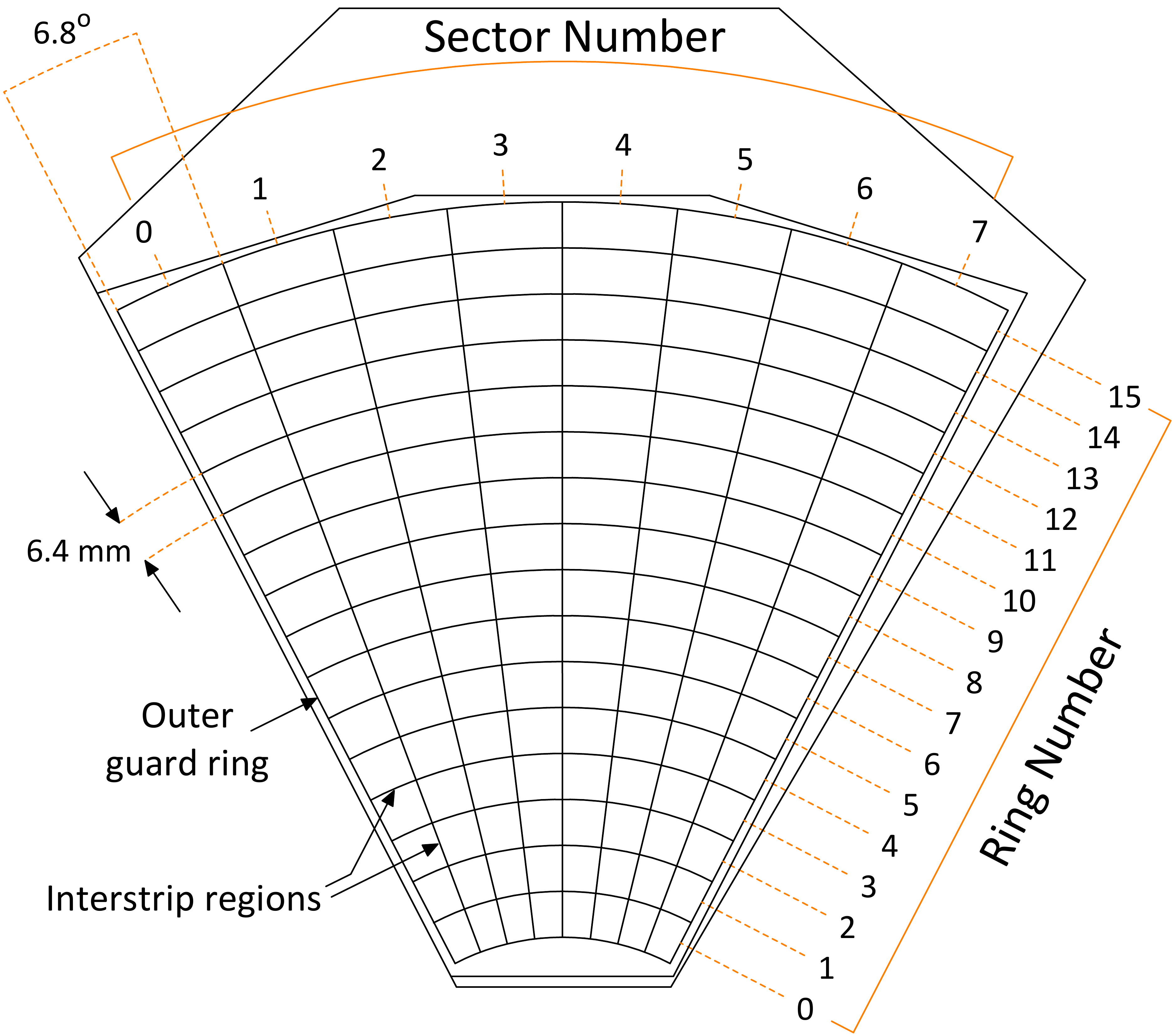}
 \caption{MMM detector from Micron Semiconductor Limited \cite{KCWL_MSc}. }
 \label{fig:MMM_detector}
\end{figure}

The CAKE is comprised of up to five double-sided silicon strip detectors (DSSSDs) of the MMM design from Micron Semiconductor Limited (Figure \ref{fig:MMM_detector}). The detectors are 400-$\mu$m thick  and have 16 ring channels on the junction side of the detector and 8 sector channels on the ohmic side. The MMM design is a fairly large (active area of 54000 mm$^2$: 102.5 mm in length covering an azimuthal angle of 54\textdegree\ when placed normal to the beam direction) wedge-shaped silicon detector that can be packed together to form a lampshade configuration, similar to that of the Yale Lampshade Silicon Array, YLSA \cite{YLSA}. Each detector covers approximately 5\% solid angle meaning that CAKE is 25\% efficient. In theory, the forward angles may also be instrumented with silicon detectors giving a total efficiency of 50\%. However, the rate of scattered beam at forward angles is too high for this to be practical. The angular range covered is approximately 115\textdegree-165\textdegree\ with angular resolution of 4\textdegree. 

An array of DSSSDs is preferred over an array of resistive-strip silicon detectors (RSSDs) because of the variations in positional resolution in RSSDs, due to variations in the deposited energy, and because these detectors can have a position-dependent threshold. This contravenes one of the experimental requirements of the CAKE. Additionally, with the use of DSSSDs, by implementing the condition that the energies measured in the front and back of the silicon detector for an event should be approximately the same, the rate of false coincidences and therefore the background in the coincidence spectra can be reduced.

\subsection{Associated mechanical infrastructure}
\label{sec:AMI}

\begin{figure}
\centering
\includegraphics[width=\textwidth]{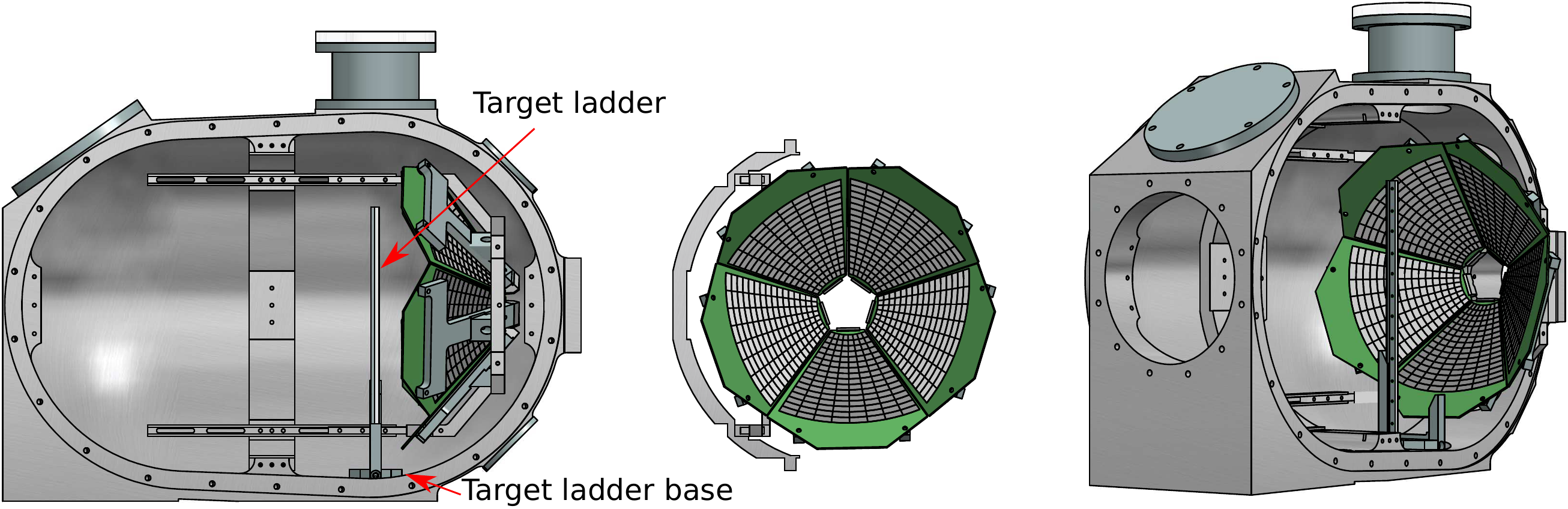}
 \caption{A diagram showing how the CAKE is mounted within the new K600 scattering chamber. (Left) A view of the CAKE from beam left, the beam enters from the right side of the diagram. The target ladder and the base are marked to show how the base of the ladder shadows the lower MMM detectors when the target ladder is in the final position. (Centre) A view looking upstream showing the CAKE lampshade geometry. (Right) A view showing in another perspective of how the CAKE fits into the scattering chamber.}
 \label{fig:CAKE_in_chamber}
\end{figure}

The CAKE and its associated infrastructure comes as part of a wider programme expanding the coincidence capabilities of the K600 magnetic spectrometer with arrays of silicon detectors and the AFRODITE array of high-purity germanium clover detectors \cite{AFRODITE} (a project entitled \textquoteleft BaGeL\textquoteright) \cite{BAGEL_Talk}. As part of this effort, a new scattering chamber - shown in Figure \ref{fig:CAKE_in_chamber} - for the K600 has been designed and constructed. This chamber has thin aluminium walls to limit attenuation of $\gamma$ rays and to allow the AFRODITE clovers to be packed efficiently around the target position.

The CAKE and the new scattering chamber were designed together and the CAKE sits at backward angles within the scattering chamber, as shown in Figure \ref{fig:CAKE_in_chamber}. The chamber was machined at the Rapid Product Development Laboratory at Stellenbosch University \cite{RPD}. The detectors comprising the CAKE have a minimum distance from the target to the detector of 100 mm and a maximum distance of 110 mm; the target-detector distance is dependent on the position on the detector.

The CAKE is designed so that the target ladder (see Figure \ref{fig:CAKE_in_chamber}) for the K600 has unimpeded movement across the open side 
of the array, with around 1 cm of clearance between the detectors and associated cables and the ladder. The ladder base is quite broad and shadows parts of the lowest two silicon detectors when the lowest target ladder position is used. However, an empty frame is always included on the target ladder to allow the beam halo to be monitored during the experiment and is always positioned in the bottom target position.

Before an experiment, each MMM detector comprising the CAKE is mounted separately onto the supporting frame. The supporting frame and the individual mounts for the silicon detectors are shown in Figure \ref{fig:CAKE_in_chamber}. The largest part of this frame is a rib, clearly visible in Figure \ref{fig:CAKE_in_chamber}, which is designed to fit onto the mounting points within the scattering chamber. To install the CAKE in the scattering chamber, the entire array can be picked up by the rib and secured in the chamber. This process is relatively quick, resulting in a total of approximately 2 hours of downtime including the time taken to vent the chamber and pump back down. During the initial beam-tuning and dispersion-matching procedure for K600 0-degree experiments, it is necessary to spend a considerable amount of time observing the beam-spot shape and size using viewers before the scattering chamber and at the target position. These viewers comprise thick zinc sulphide scintillators, increasing both the amount of back-scattered beam seen by the CAKE and the radiation damage to the silicon detectors. As the CAKE may be installed quickly in the chamber, the initial beam tuning can be done without the silicon detectors in the chamber and so the exposure to scattered beam during the initial tuning stages of the experiment is minimised. Once the beam is optimised, the CAKE is installed. This satisfies the final design specification (6) of the CAKE.

The K600 in 0\textdegree-mode is extremely sensitive to scattered beam and to beam halo. If additional scattering points are introduced downstream of the target, then elastically scattered beam from the target can be re-scattered into the spectrometer, causing a high background rate at the focal plane. With the CAKE installed, the only additional material close to the beam is from the thin ends of the MMM detectors (which form a pentagon of around 2 cm per side) and some material for its support structure. However, the aperture formed from the MMM detectors is larger than the target frame and is therefore unlikely to cause additional scattering if the beam is well-tuned. In addition, it is upstream of the target position and thus cannot re-scatter elastically scattered beam particles into the spectrometer.

Finally, in order to operate the CAKE and the K600 together, the signals from the CAKE must be fed into the K600 data acquisition (DAQ). The in-vacuum cables were manufactured specifically for the CAKE: the 50-pin connector on the MMM detector is split into two cables, both of which are shielded by metal braid. The signals are then fed into 19-pin LEMO feedthroughs. This means that bulky breakout boards which would not easily fit within the compact scattering chamber do not have to be used. Outside the chamber, standard Mesytec braided cables are used to carry the signals from the LEMO feedthroughs to the preamplifiers.

\subsection{Electronics and data acquisition}

The CAKE uses Mesytec preamplifiers (models MPR-16 and MPR-32) for initial signal amplification. Signals are then fed into Mesytec MSCF-16 constant-fraction discriminator amplifiers which give shaped energy output signals that are fed into the CAEN V785 analogue-to-digital converters (ADCs) in the VME-based K600 DAQ. The timing signals from the CAKE are fed into the same V1190A TDCs as used for the K600 timing information. These TDCs store the timing information on a rolling buffer. After a trigger signal is received by the TDC module, the DAQ records events stored on the rolling buffer from up to 800 ns backwards in time and 1200 ns forwards in time. This negates the requirement that the silicon signals be delayed in order to fall after the focal plane trigger.

The experimental trigger is given by the focal-plane plastic scintillating paddle(s) and, in addition to triggering the data acquisition also generates a $6-\mu s$-wide ADC gate signal which is fed into the V785 modules. The signal from the CAKE given to the ADCs must fall within this gate taking into account the processing time and delays in the various electronics. In order to ensure that this is the case, the shaping time of the MSCF-16 amplifiers is chosen to be 1 $\mu$s. This effectively delays the signal so that the peak of the shaper output falls comfortably within the ADC gate.

\section{Performance of the CAKE}
\label{sec:perfomance}

\begin{figure}
\includegraphics[width=\textwidth]{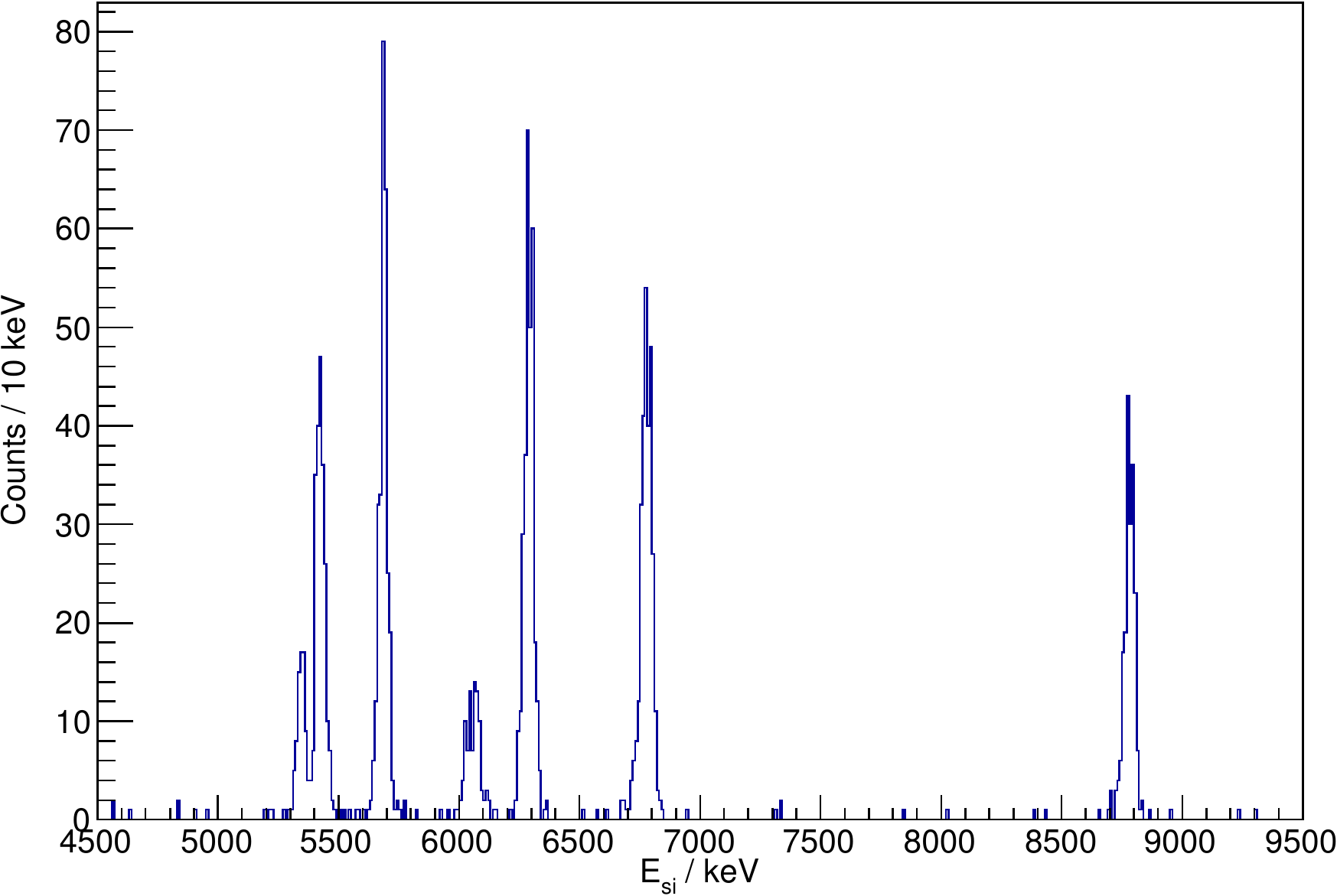}
 \caption{A calibrated $\alpha$-particle spectrum from a $^{228}$Th source for a single junction strip on an MMM detector. The energy resolution is 45 keV, FWHM.}
 \label{fig:MMM_alpha_spectrum}
\end{figure}

The CAKE is typically calibrated for its energy response using a $^{228}$Th source. The energy resolution achieved for each strip is around 40-50 keV, FWHM (see Figure \ref{fig:MMM_alpha_spectrum}). Degradation of the energy resolution of some of the silicon detectors has been observed after experiments, particularly for the ohmic sides of the detectors. This has been observed by comparing the pre-experiment $^{228}$Th source calibration runs to the post-experiment calibration runs. In these cases, the energy measured in the ohmic side of the detector is slightly below (20-40 keV) the full-energy peak when the pre-experiment calibration is used. For rings that are further forward (i.e. the outer rings which are at lower $\theta$) the effect is more pronounced suggesting that it results from scattered beam causing damage to the detector.

The CAKE is able to detect charged particle with energies above 400 keV, the primary limitation on this is from beam-induced background such as elastically scattered beam particles hitting the silicon detectors. For practical purposes, the trigger rate in the silicon detectors is limited to around 15 kHz, corresponding to around 1 kHz per strip in the detector which is around the maximum rate per strip in a silicon detector. In experiments using the $(p,t)$ reaction on $^{50}$Cr and $^{24}$Mg, beam intensities of up to 20 pnA were used.

It should be noted that the beam current limitation of the K600 and the CAKE is highly dependent on a number of factors including the reaction of interest, the target thickness and composition and the beam energy. For inelastic scattering experiments, the focal plane rate is nearly always the limitation, especially for targets with high atomic number. For transfer reactions where higher beam currents are required to collect the required amount of statistics, then the limitation on the beam current is the focal-plane energy resolution due to the dispersion-matching of the accelerator system \cite{Neveling}.

For protons, there is a limitation on the highest energy which can be detected of 7 MeV, the energy at which the protons punch through the 400-$\mu$m thick MMM detectors. The lowest measureable energy is limited by the electronics to around 400 keV. For $\alpha$ particles, the lower limit on the detectable energy is also around 400 keV with an upper limit of around 28 MeV before punch through. For any decaying species, however, the performance of the detector and electronics is not usually the limiting factor on the lowest detectable energy. Rather, the target thickness attenuating the decay particle is often the decisive quantity. This depends strongly on the target composition and thickness and thus varies strongly between experiments. Note that the 400-keV lower limit is on the measured energy; a thick target which causes considerable attenuation of a low-energy decay particle may mean that this particle falls below the 400-keV threshold.

\begin{figure}
 \includegraphics[width=\textwidth]{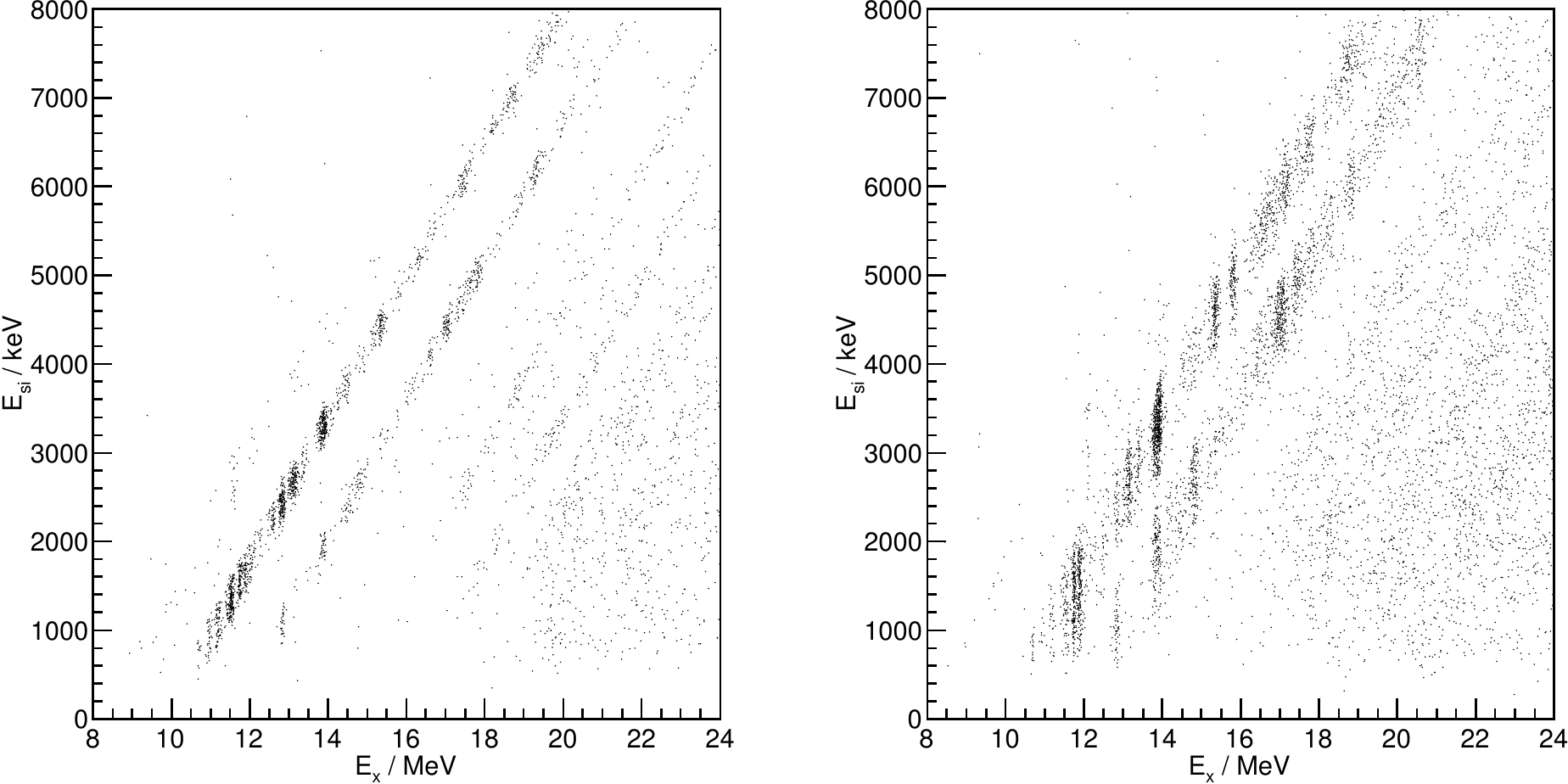}
 \caption{Two coincidence spectra from (left) an inner ring at high $\theta$ and (right) an outer ring at lower $\theta$. The effect of the additional target thickness for the ring at a shallower angle is evident in the larger vertical spread of the coincidence loci in the right-hand spectrum compared to the left-hand one. The data are from a $^{24}$Mg($\alpha,\alpha^\prime$)$^{24}$Mg($\alpha$)$^{20}$Ne coincidence experiment using a 200-MeV $\alpha$-particle beam and a $230$-$\mu$g/cm$^2$-thick $^{24}$Mg target.}
 \label{fig:target_thickness_effect}
\end{figure}

In many experiments the limitation on the energy resolution in the CAKE is not the intrinsic resolution of the silicon detectors or of the electronics but is rather the relatively thick targets used in K600 experiments. This effect is demonstrated very clearly in Figure \ref{fig:target_thickness_effect}: the left-hand panel is from a strip at the most backward angles and the right-hand panel is from a strip at a smaller angle. The coincidence energy loci are clearly wider in the right-hand panel compared to the left-hand panel due to the change in effective target thickness. Taking as an example the $\alpha_0$ decay ($E_\alpha = 3.81$ MeV) from the two $0^+$ states at 13.89 MeV in $^{24}$Mg, the energy resolution of the most backward (forward) strip of the CAKE is 275 (725) keV, FWHM. For the proton decays from the same states ($E_p = 2.11$ MeV), the resolution in the most backward (forward) strip if CAKE is 125 (290) keV, FWHM. 

\begin{figure}
 \includegraphics[width=\textwidth]{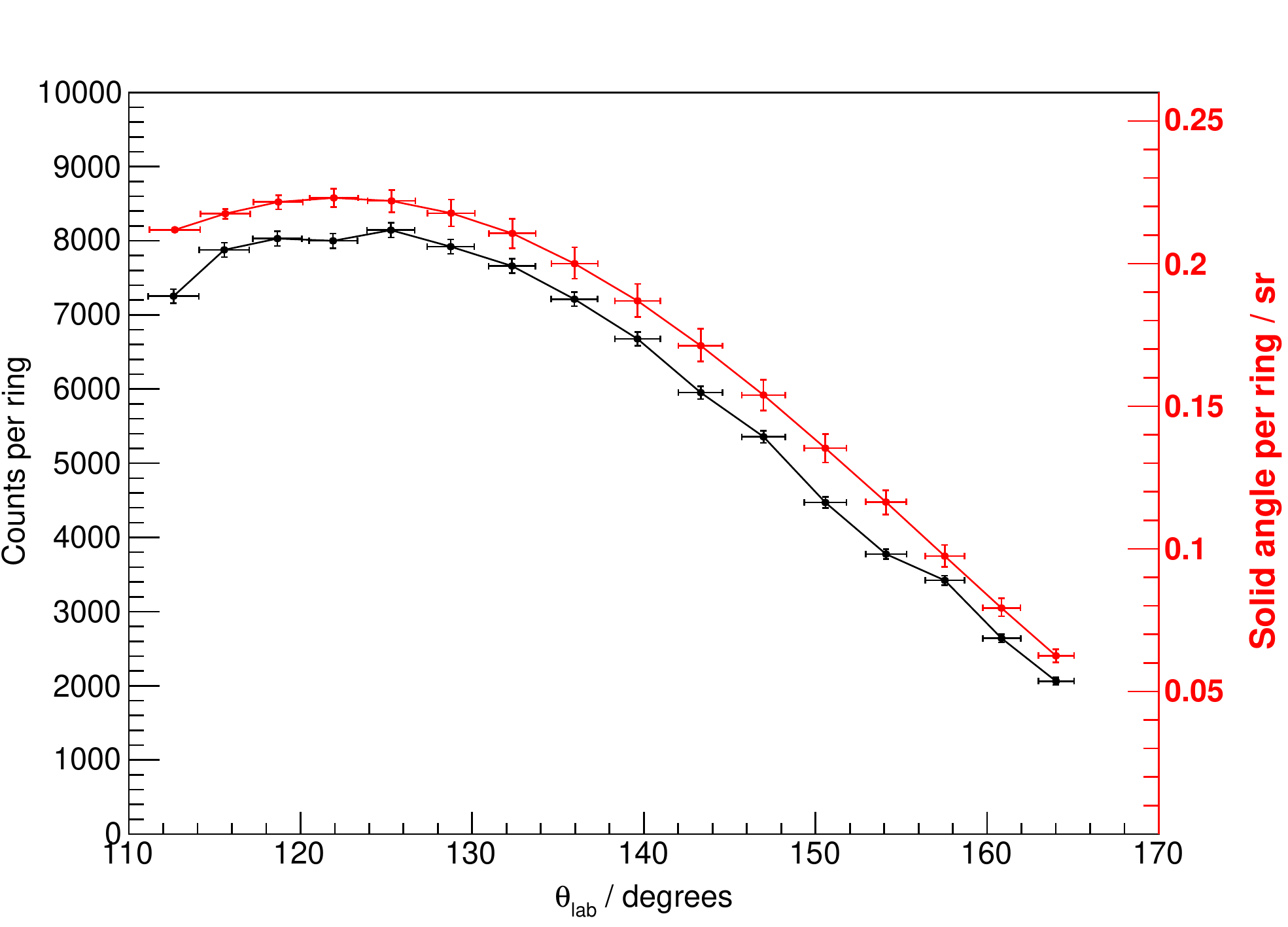}
 \caption{The counts per ring for decay from a $J^\pi=0^+$ state (black, left axis) compared to the simulated solid angle per ring (red, right axis). The data and the simulation are from Refs. \cite{KCWL_MSc,KCWL_arXiv} in which only four MMM detectors were used. The horizontal ($\theta$) error bars, for both the simulation and the experimental data, depict one standard deviation of the angle distribution subtended by each ring. The vertical error bar for the data is purely statistical, whilst for the simulated solid angle, the error represents the range in solid angle corresponding to a $\pm2$ mm position error of the CAKE along the beam axis.}
 \label{fig:KCWL_source_data}
\end{figure}

The solid angle of the CAKE can be calculated using GEANT4 simulations and compared to calibration source data for which the source is placed at the target position of the K600 \cite{KCWL_MSc}. The simulated solid angle and the measured counts from an isotropically decaying state are shown in Figure \ref{fig:KCWL_source_data} with good agreement between the simulation and the data.

\begin{figure}
  \includegraphics[width=\textwidth]{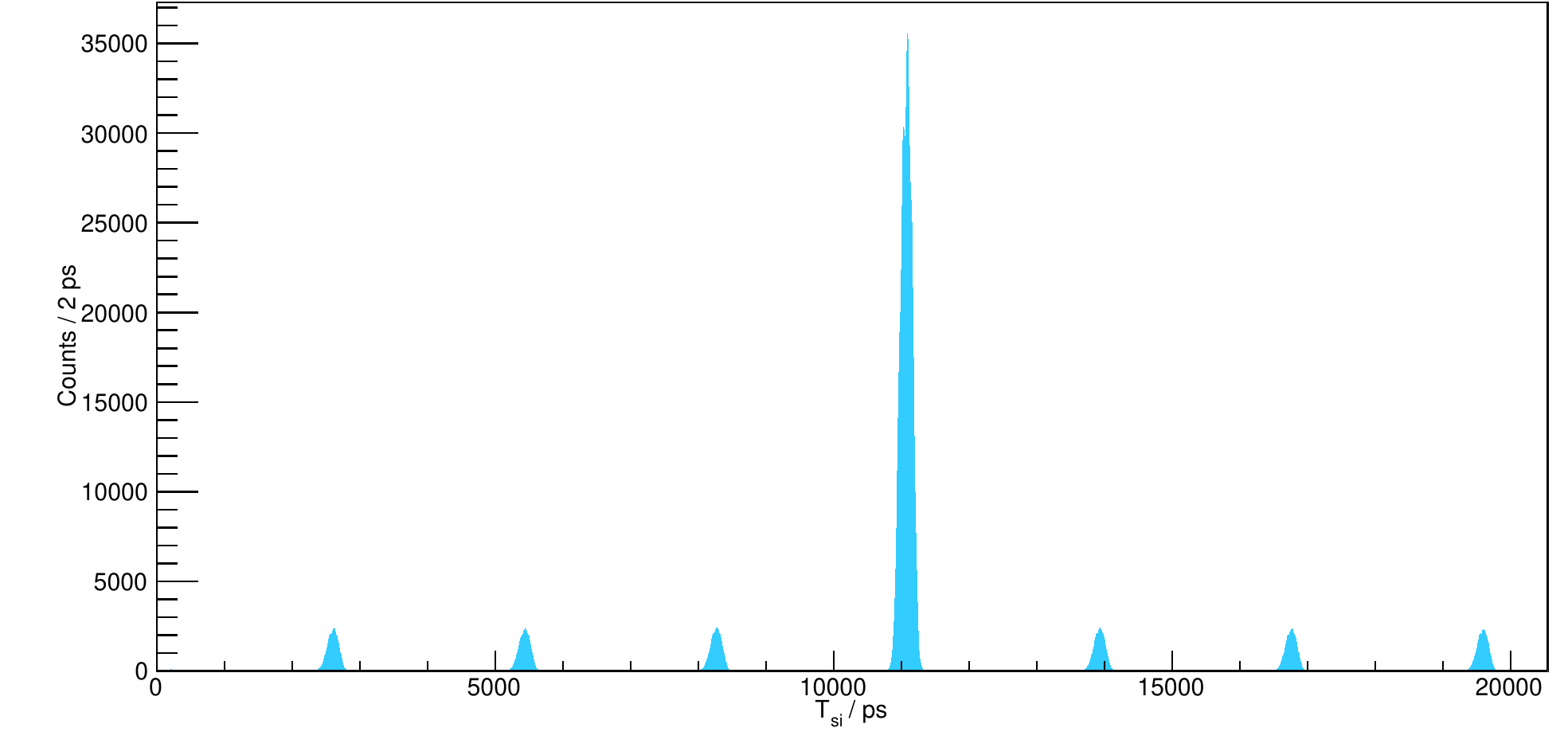}
 \caption{CAKE raw timing spectrum from the $^{16}$O($\alpha,\alpha^\prime$)$^{16}$O experiment \cite{KCWL_MSc,KCWL_arXiv}. This experiment was performed using pulse selection; only 1 in 5 pulses were selected for use, with a separation between pulses of 2 ns. The real event-to-random event ratio is 14.7}
 \label{fig:CAKE_Raw_Timing}
\end{figure}

Timing information is an important consideration when using the CAKE. First of all, the real K600+CAKE coincidences must have a well-defined timing separation. In Figure \ref{fig:CAKE_Raw_Timing}, an example timing spectrum for the CAKE is shown, the main pulse is the real coincidence peak and the other peaks are due to random coincidences between a focal plane event and a silicon detector event. If pulse selection (see Section \ref{sec:K600}) is not utilised during the experiment, it becomes more difficult to firmly assign which focal plane hit corresponds to which CAKE event as a number of beam pulses can fall within the time window. Pulse selection separates the beam pulses sufficiently for clear associations to be made. It is important to note that this is an additional consideration to the use of the silicon timing for particle identification as described below: if it is not possible to make a firm association between the focal plane hit and the silicon detector hit, it is not possible to make a clean particle identification gate in the CAKE.

The second use of timing information is to identify the particle hitting the CAKE. This is done by considering the time between the silicon detector hit and the next RF signal (which signifies the subsequent beam pulse). Due to the details as to how the data are recorded in the K600 DAQ, this information is obtained by considering the time difference between the silicon hit and the K600 spectrometer time-of-flight (for details, consult Ref. \cite{Neveling}). The final timing resolution for the silicon time-of-flight is around 2.5 ns, FWHM. This includes the intrinsic resolution of the silicon detectors, the timing resolution of the focal plane plastic paddle and the RF signal from the accelerator and the timing resolution of the silicon detector electronics.

\begin{figure}
\includegraphics[width=\textwidth]{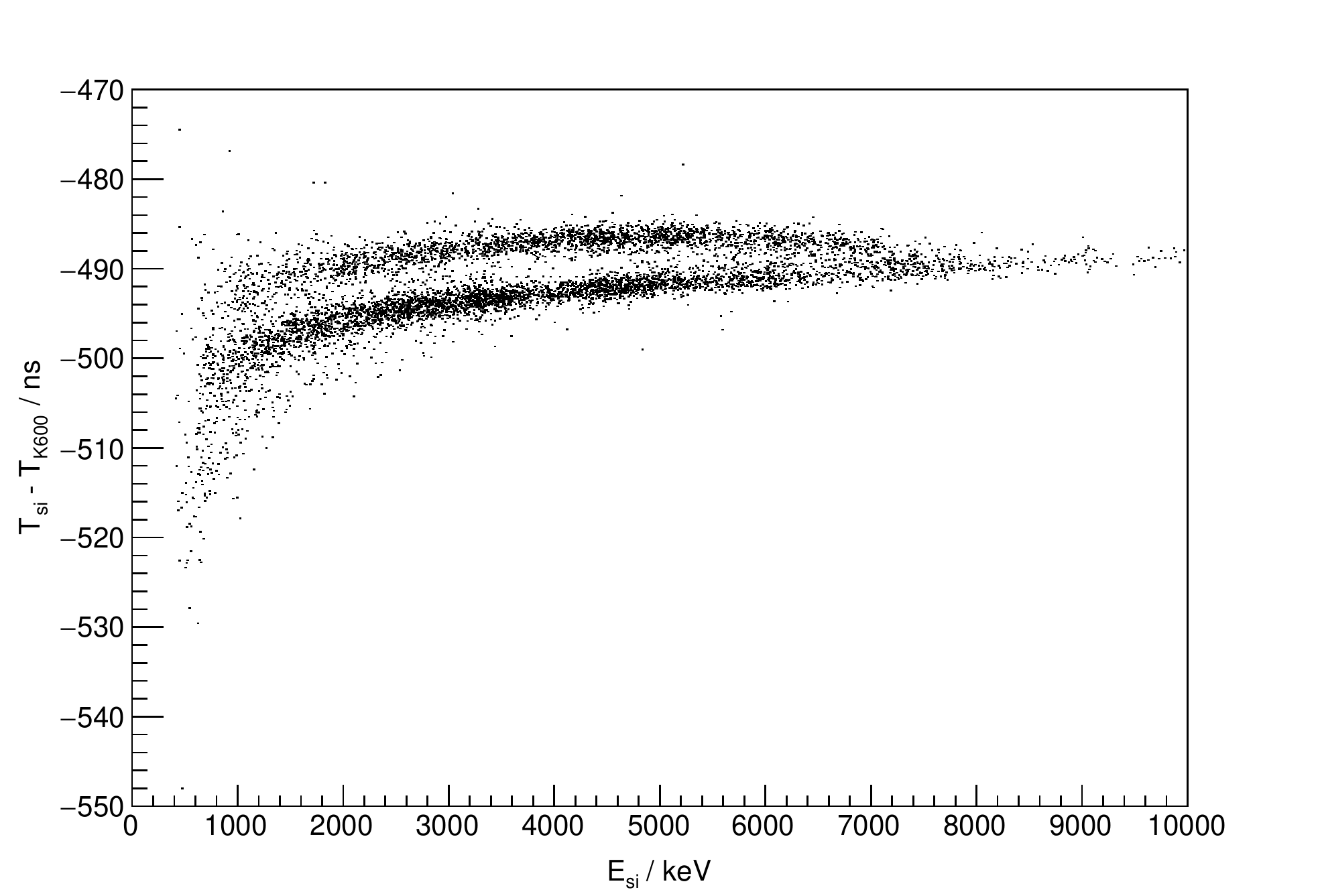}
 \caption{Relative time-of-flight to the CAKE vs silicon energy loci. The upper locus, terminating at around 7000 keV represents the observed protons and the lower locus represents $\alpha$ particles.}
 \label{fig:CAKE_PID_through_timing}
\end{figure}

One complication with the PID loci is the shape of the proton locus. Protons above about 7 MeV will punch through the silicon detector. Before this happens, the timing signal from the silicon detector shifts due to changes in the pulse shape from the silicon detectors resulting in a region of overlap in the CAKE PID spectrum around 7 MeV, as shown in Figure \ref{fig:CAKE_PID_through_timing}.

\begin{figure}
\includegraphics[width=\textwidth]{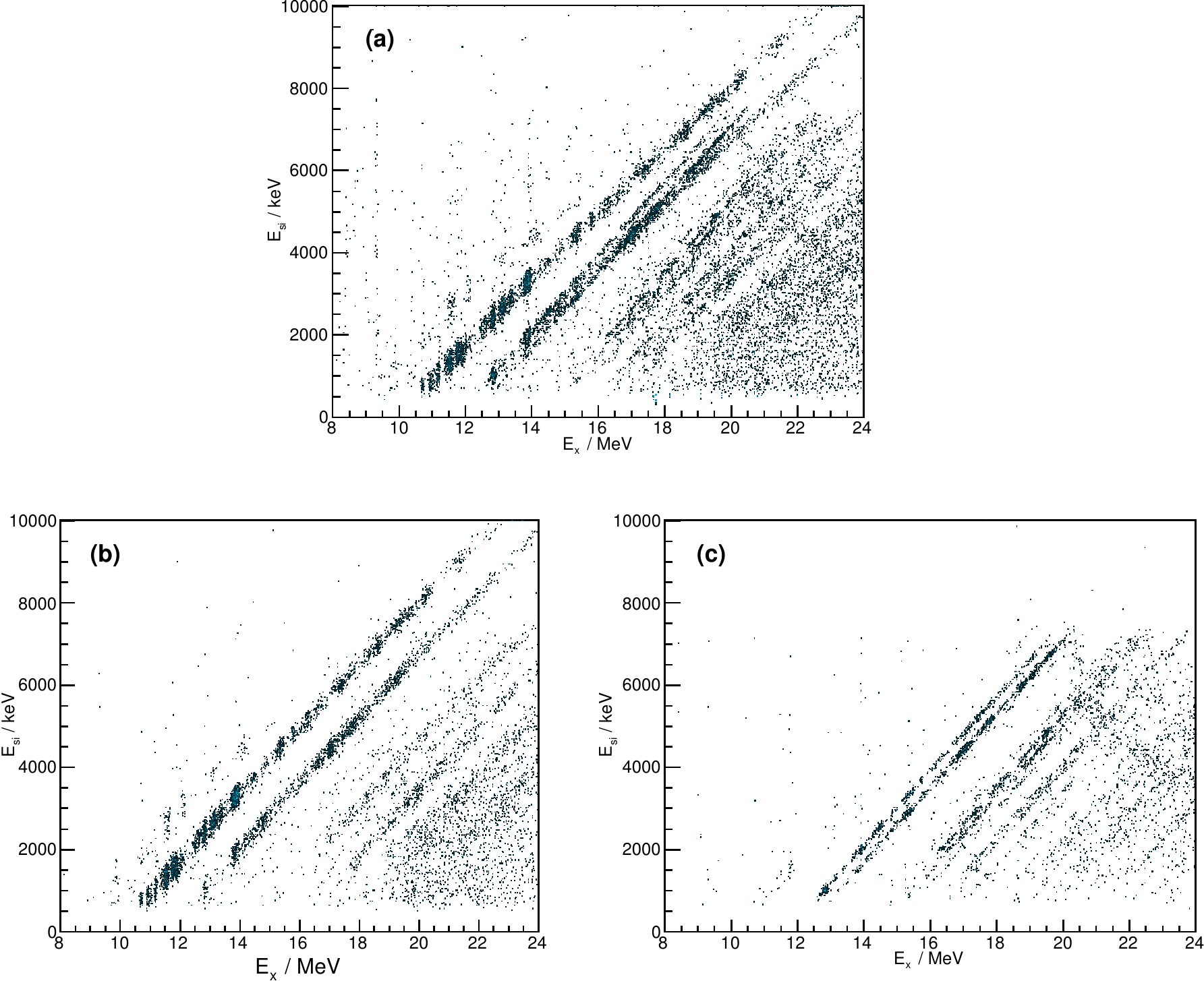}
 \caption{Coincidence spectra from the $^{24}$Mg($\alpha,\alpha^\prime$)$^{24}$Mg reaction for (top) coincidence events without a timing gate imposed, (lower left) $\alpha$-particle decays selected in the CAKE and (lower right) proton decays.}
 \label{fig:2D_spectrum}
\end{figure}

Using the silicon time-of-flight vs silicon energy loci to select different decay channels, excitation energy vs silicon energy matrices can be constructed, showing how different excited states decay. The ungated 2D spectrum is shown in Figure \ref{fig:2D_spectrum}(a) and the 2D spectra gated on $\alpha$ particles and protons are shown in Figure \ref{fig:2D_spectrum}(b) and (c) respectively. The effect of proton punchthrough of the detectors can be seen in Figure \ref{fig:2D_spectrum}(c): in addition to the diagonal loci representing decays to various states in $^{23}$Na, there is a locus cutting across almost perpendicular to these decay loci. This additional locus is caused by proton punchthrough from decays to the ground and first-excited states in $^{23}$Na.

\section{Future Developments}

In contrast to traditional scattering chambers for magnetic spectrometers, the new K600 scattering chamber does not have a sliding seal which means that presently it can only be used at 0 degrees and 4 degrees. Modifications to the new scattering chamber are underway which will allow the spectrometer to measure at angles from 0\textdegree\ to 39\textdegree\ with a blind spot from 14\textdegree\ to 
17\textdegree\ caused by the support frame of the chamber. This will allow for coincidence measurements to be made at multiple angles. However, at non-zero angles below 20\textdegree, unscattered beam cannot be transported away to a remote beam-stop and must therefore be stopped in front of the spectrometer quadrupole, increasing the background in any ancillary detector systems around the target position, though previous experimental studies using magnetic spectrometers and silicon detectors have managed to limit this problem by shielding the silicon detectors from the beamstop \cite{Orsay}. 

\section{Summary}
\label{sec:summary}

The Coincidence Array for K600 Experiments at iThemba LABS is a new ancillary detector system for use with the K600 magnetic spectrometer. It provides scientists with a valuable new tool for the study of charged-particle decays of excited states in nuclei for science cases involving nuclear structure and nuclear astrophysics. The particle discrimination capabilities, low background conditions as well as the high overall efficiency of the CAKE represents a meaningful improvement compared to similar detector system used elsewhere. The combination of such an array with the capability of the K600 magnetic spectrometer to measure inelastic scattering and pickup reactions at extreme forward angles, including zero degrees and with the intermediate-energy beams available from the Separated Sector Cyclotron at iThemba LABS, results in a globally unique facility.

The experimental programme for the K600 and the CAKE is underway. It is anticipated that a new cyclotron will be acquired for medical radio-isotope production at iThemba LABS. This will lift the isotope production load from the K = 200 separated-sector cyclotron at iThemba LABS, which will allow more time to be dedicated to nuclear physics research. It is expected that experiments using the K600 and the CAKE will form a significant portion of the nuclear physics research at iThemba. We anticipate programmatic studies of, for example, the $\alpha p$-process in Type I X-ray bursts could take place with the
increased beamtime available.

\acknowledgments

The authors thank the accelerator group at iThemba LABS for the provision of a variety of high-quality halo-free dispersion-matched beams. The K600 is supported by the NRF and the CAKE was funded by the NRF under grant number 86052. RN acknowledges financial support from the NRF through grant number 85509. We thank the University of York for loaning additional electronics for the silicon detectors and the K600 DAQ. PA thanks Michael Munch and Alan Howard of Aarhus University for advice on the instrumentation of the silicon detectors.

\end{document}